\documentclass[twocolumn,showpacs,amsmath,amssymb,aps]{revtex4}
\newtheorem{lemma}{Lemma}

\newtheorem{theorem}{Theorem}
\def\<{\langle}\def\>{\rangle}\def\Tr{\operatorname{Tr}}
\def\map#1{{\mathcal{#1}}}\def\set#1{{\sf #1}}
\def\dual#1{{#1}^\tau}\def\rnk{\operatorname{rank}}
\def\Rng{\set{Rng}}\def\dim{\operatorname{dim}}
\def\sH{\set{H}}\def\sK{\set{K}}\def\Bnd#1{\set{B}(\set{#1})}
\def\T#1{\set{T}(\set{#1})}\def\sD{\set{D}}\def\sS{\set{S}}\def\sR{\set{R}}
\def\sL{\set{L}}\def\Proof{\medskip\par\noindent{\bf Proof. }}
\def\qed{$\blacksquare$\par}\def\poper{\operatorname{+}}
\def\hplus{\sideset{{\cdot\,}}{\cdot}\poper}
\def\vplus{\overset{{\cdot}}{\underset{\cdot}{\poper}}}
\def\kk{\rangle\!\rangle}\def\bb{\langle\!\langle}
\begin{document}
\title{Physical realizations of quantum operations} \author{Francesco
Buscemi} \email{buscemi@fisicavolta.unipv.it} \author{G.  Mauro
D'Ariano} \email{dariano@unipv.it} \altaffiliation[Also at
]{Department of Electrical and Computer Engineering, Northwestern
University, Evanston, IL 60208} \author{Massimiliano F.  Sacchi}
\email{msacchi@unipv.it} \affiliation{Quantum Optics \& Information
Group, INFM Unit\`a di Pavia} \affiliation{Universit\`a di Pavia,
Dip.to di Fisica ``A. Volta'', via Bassi 6, I-27100 Pavia, Italy}
\homepage{http://www.qubit.it} 
\date{\today}% It is always \today, today,
\begin{abstract}
  Quantum operations (QO) describe any state change allowed in quantum
  mechanics, such as the evolution of an open system or the state change
  due to a measurement.  We address the problem of which unitary
  transformations and which observables can be used to achieve a QO
  with generally different input and output Hilbert spaces. We
  classify all unitary extensions of a QO, and give explicit
  realizations in terms of free-evolution direct-sum dilations and
  interacting tensor-product dilations. In terms of Hilbert space
  dimensionality the free-evolution dilations minimize the physical
  resources needed to realize the QO, and for this case we provide
  bounds for the dimension of the ancilla space versus the rank of the
  QO. The interacting dilations, on the other hand, correspond to
  the customary ancilla-system interaction realization, and for these
  we derive a majorization relation which 
  selects the allowed unitary interactions between system and ancilla.
\end{abstract}
\maketitle
\section{Introduction}
The recent progresses in quantum information theory
\cite{pop,Nielsen,sv} offer the possibility of radically new
information-processing methods that can achieve much higher
performances than those obtained by classical means, in terms of
security, capacity, and efficiency \cite{cry,ben,bra,ben2}. This urges
a quantum system engineering approach for the production of the new
quantum tools needed for communication, processing, and storage of
quantum information. A first step toward this goal is the search for a
systematic method to implement in a controlled way any quantum state
transformation.  \par The mathematical structure that describes the
most general state change in quantum mechanics is the {\em quantum
operation} (QO) of Kraus \cite{Kraus83a,Nielsen}. Such abstract
theoretical evolution has a precise physical counterpart in its
implementations as a unitary interaction between the system undergoing
the QO and a part of the apparatus---the so-called {\em
ancilla}---which after the interaction is read by means of a
conventional quantum measurement.  In this paper we address the
problem of which unitary transformations and which observables can be
used to achieve a given QO for a finite dimensional quantum system. We
consider generally different input and output Hilbert spaces $\sH$ and
$\sK$, respectively, allowing the treatment of very general quantum
machines, e. g. of the kind of quantum optimal cloners
\cite{wern,clon}. As it will be clear from the physical
implementations of the QO, schematically this corresponds to the
general scenario, in which the machine prepares a state in the Hilbert
space $\sH$ and couples it unitarily with a \emph{preparation} ancilla
in the Hilbert space $\sR$, which was previously set to a fixed
state. The machine then transfers the joint system with Hilbert space
$\sR\otimes\sH$ to a measuring section, which performs a measurement
on another \emph{measurement} ancilla, with space $\sL\subset\sR
\otimes\sH$. The output system will be in the Hilbert space $\sK$,
where $\sK$ is such that $\sL\otimes\sK=\sR\otimes\sH$. The result is
a machine that performs a QO with input in $\sH$ and output in $\sK$.
\par In the process of classification of all unitary extensions of a
QO, we will give explicit realization schemes in terms of
free-evolution direct-sum dilations and interacting tensor-product
dilations, which in the following will be named shortly {\em free} and
{\em interacting} dilations, respectively. The interacting dilations
correspond to the ancilla-system interaction scenario just described
above, whereas in the {\em free} dilations we have only the
measurement ancilla, and the input space is embedded in a larger
Hilbert space, where a kind of super-selection rule forces the choice
of the input state in a proper subspace before a free unitary
evolution.  In terms of Hilbert space dimensionality the free
dilations minimize the physical resources needed to realize the QO,
and for this case we will provide bounds for the dimension of the
ancilla space versus the rank of the QO. For the interacting
dilations, on the other hand, we will derive a majorization relation
which allows to pre-select the admissible unitary interactions between
system and ancilla, in relation with the ancilla preparation state and
the measured observable.  \par The paper is organized as
follows. After briefly recalling the notion of quantum operation in
Sec. II, in Sec. III we introduce the Stinespring form for a QO and
explicitly construct all possible unitary realizations, for both free
and interacting dilations.  We also address the problem of finding
{\em unitary interacting power dilations} of a given QO, namely
interacting dilations that also provide the $k$-th power of the map
(i.e. with the map applied $k$ times). In Sec. IV we give the
criterion to select the admissible unitary interactions for a QO in
form of a majorization relation. Section V finally closes the paper
with a summary of the results.
\section{Quantum operations} In the following by $\T{H}$ we
denote the set of trace-class operators on the Hilbert space $\sH$
(which can be simply regarded as just the set of states on $\sH$). A
{\em quantum operation\/} $\map{E}:\T{H}\rightarrow\T{K}$ is a
linear, trace not-increasing map that is also \emph{completely
  positive} (CP), namely that preserves positivity of any input state
of the system on $\sH$ entangled with any other quantum system
(mathematically, all trivial extensions $\map{E}\otimes\map{I}$ of the
map must preserve positivity of input states on the extended Hilbert
space). The \emph{input} and the \emph{output} states are connected
via the relation
\begin{equation}\label{start}
\rho\longmapsto\rho'=\frac{\map{E}(\rho)}{\Tr[\map{E}(\rho)]}\;,
\end{equation}
where the trace $\Tr[\map{E}(\rho)]\le 1$ also represents the
probability that the transformation in Eq. (\ref{start}) occurs.
An analogous of the spectral theorem for positive operators in finite
dimensions leads to the following canonical form of the QO  
$\map{E}:\T{H}\rightarrow\T{K}$
\cite{Kraus83a}
\begin{equation}\label{kraus-mix}
{\cal E}(\rho) = \sum_n E_n \rho E_n^{\dag}\;, 
\end{equation}
where the bounded operators $E_n\in\Bnd{H,K}$ from $\sH$ to
$\sK$ are orthogonal, i.e. $\Tr[E_n^\dag E_m]=0$ for $n\neq m$, and
moreover they satisfy the condition
\begin{equation}\label{sum}
\sum_n E_n^\dagger E_n=K\le I_{\sH}\;.
\end{equation}
In terms of the positive operator $K\in\Bnd{H}$, the probability of
occurrence of the QO can also be rewritten as $\Tr[K\rho]$ . Notice
that there are generally infinitely many non-canonical ways of writing
the map $\map{E}$ in the form of Eq.  (\ref{kraus-mix}), with
generally larger and non orthogonal sets of elements $\{E'_j\}$ that
satisfy Eq. (\ref{sum}). All such decompositions are usually called
{\em Kraus forms} of the QO ${\cal E}$. In order to satisfy Eq.
(\ref{sum}), the operators $\{E'_j\}$ of a non-canonical Kraus form
are related to the canonical ones $\{E_i\}$ as $E'_j=\sum_iY_{ji}E_i$
via an isometric matrix $Y$, i.e.  a matrix with orthonormal columns.
When the map is trace-preserving, i.e. $\Tr[\map{E}(\rho)]=1$---or
equivalently $K=I_\sH$---it occurs with unit probability, and is
usually named {\em channel}.

\par It was known since Kraus \cite{Kraus83a} that a trace-preserving QO
admits a unitary realization on an extended Hilbert space. More
generally, when we have a set of QO's that describe a general quantum
measurement (also with continuous spectrum and in infinite dimensions:
the so-called \emph{instruments}) Ozawa \cite{Ozawa84}
proved the realizability in terms of an observable measurement over an ancilla after
a unitary interaction with the quantum system. In the following we
will derive explicitly all possible unitary dilations for a generic QO
for finite dimension, and give ancillary realizations and bounds for
the dimensions of the involved Hilbert spaces.

\section{Unitary dilations of a quantum operation}
The Stinespring dilation \cite{Stinespring,EvansLewis} is a kind of
``purification'' of the QO. Originally, the Stinespring's theorem was
set for the \emph{dual} version $\dual{\map{E}}$ of the QO, i.e. in
the ``Heisenberg picture''---instead of the ``Schr\"odinger picture'' of
Eq. (\ref{start})---the two pictures being related as follows
\begin{equation}
\Tr[\rho\;\dual{\map{E}}(O)]=\Tr[\map{E}(\rho)\;O]\;,\label{dual}
\end{equation}
for every bounded operator $O\in\Bnd{K}$. Analogously to Eqs. 
(\ref{kraus-mix}) and (\ref{sum}), one has 
\begin{equation}
\dual{\map{E}}(O)=\sum_nE_n^\dag OE_n\label{dualmap}
\end{equation}
with
\begin{equation}
\dual{\map{E}}(I_{\sK})=K.
\end{equation}
\par A variation of the
Stinespring's theorem can be restated by saying that for every QO
$\map{E}:\T{H}\rightarrow\T{K}$, there exists a
Hilbert space $\sL$ such that $\map{E}$ can be obtained as follows
\begin{equation} \label{scarpa}
\dual{\map{E}}(X)=E^\dag(I_\sL\otimes X)E\;,
\end{equation}
where $E\in\Bnd{H,L\otimes K}$ is a contraction (i.e. $E$
is an operator bounded as $\|E\|\leq 1$).  In fact, consider any Kraus
decomposition $\map{E}=\sum_{i=1}^n E_i\cdot E_i^\dag$ for $\map{E}$,
and let $\sL$ be a Hilbert space with $\dim(\sL)\ge n$ and orthonormal
basis $\{|l_i\>\}$.  The following operator  
\begin{equation}
E=\sum_{i=1}^{\dim(\sL)} |l_i\>\otimes E_i \label{cont}
\end{equation}
is a contraction, since $E^\dag E=K\le I_{\sH}$ (if one considers
$\dim(\sL)>n$, it is meant that extra null-operators  are appended to
the Kraus decomposition). Here and throughout the paper, 
for $A\in\Bnd{H,K}$ and $|v\>\in\sL$ the tensor
notation $|v\>\otimes A $ will denote the linear operator from $\sH$ to
$\sL\otimes \sK$ defined as $(|v\>\otimes A)|\phi\>\doteq
|v\>\otimes A|\phi\>$, for $|\phi\>\in\sH$,
whereas its adjoint $\<v|\otimes A^\dag$ is
the linear operator from $\sL\otimes \sK$ to $\sH$ given by
$(\<v|\otimes A^\dag)|\varphi\>\otimes|\psi\>\doteq
\<v|\varphi\> A^\dag |\psi\>$, for $|\psi\>\in\sK$ and
$|\varphi\>\in\sL$.  Using the Kronecker representation of
the tensor product \cite{kron}, the contraction $E$ in Eq. (\ref{cont}) is easily
represented by joining vertically the operators $E_i$. By substituting
Eq. (\ref{cont}) into Eq. (\ref{scarpa}) one obtains
Eq. (\ref{dualmap}), namely the statement. On the other hand, the
Schr\"odinger picture form of Eq. (\ref{dualmap}) is
\begin{equation}
\map{E}(\rho)=\Tr_\sL[E\rho E^\dag]\;.
\end{equation}
For a trace-preserving map the Stinespring contraction $E$ is actually
an isometry, since $E^\dag E=I_{\sH}$ (this case with isometric $E$ is
the original Stinespring theorem version of Eq. (\ref{scarpa})).
\par It is possible to extend also trace-decreasing maps to isometries.
For such purpose, first we prove the following lemma.
\begin{lemma} \label{para}
  For any given positive bounded operator $P\in\Bnd{H}$ and for
  every Hilbert space $\sK$, there exists a set of bounded operators
  $A_i\in\Bnd{H,K}$, $i=1,\ldots,n$, such that
\begin{equation}
P=\sum_{i=1}^n A_i^\dag A_i\;.
\end{equation}
\end{lemma}
\Proof Let $P=\sum_{i=1}^{\rnk(P)}|v_i\>\<v_i|$, where $|v_i\>\in\sH$
are the orthogonal eigenvectors of $P$, generally not normalized.  One
has two possibilities: \newline (a) $\dim(\sK)\ge\rnk(P)$: the
statement holds for $n=1$ with $P=A^\dag A$ and
$A=\sum_{i=1}^{\rnk(P)}|k_i\>\<v_i|$, with $\{|k_i\>\}$ any
orthonormal set in $\sK$. \newline (b) $\dim(\sK)<\rnk(P)$: then the
results holds with $n=\rnk(P)$ and $A_i=|\psi_i\>\<v_i|$, with
$\{|\psi_i\>\}$ any set of normalized vectors in $\sK$. \qed Notice
that in case (b) of the proof, we can suitably choose the operators
$\{A_i\}$ in order to minimize $n$ as $n=\lceil r/k\rceil$, for
$r\doteq\rnk(P)$, $k\doteq\dim(\sK)$ and $\lceil x\rceil$ denoting the
minimum integer greater or equal to $x$. These are given by the
operators
\begin{equation}
A_i=\sum_{j=1}^k |k_j\>\< v_{(i-1)k+j}|,\qquad i=1,\ldots
n\equiv\lceil r/k\rceil.
\end{equation} \par The lemma stated above can be used to prove the
following theorem. 
\begin{theorem} \label{tuno}
A linear map
  $\map{E}:\T{H}\rightarrow\T{K}$ is a QO
 if and only if its dual form can be written as
\begin{equation}
\dual{\map{E}}(X)=V^\dag (\Sigma\otimes X) V\label{iStines}
\end{equation}
for a suitable ancillary Hilbert space $\sL$, where
$V\in\Bnd{H,L\otimes K}$ is an isometry, and $\Sigma\in\Bnd{L}$
is a non-vanishing orthogonal projector on a subspace of $\sL$.
Furthermore $\Sigma\equiv I_\sL$ if and only if $\map{E}$ is
trace-preserving.
\end{theorem}
\Proof Let us  denote by $\{|\sigma_j\>\}_{j=1,...,\rnk(\Sigma)}\subset
\sL$ the eigenvectors of $\Sigma$ having unit eigenvalue. Then,
the operators
\begin{equation}
E_j=(\<\sigma_j|\otimes I_\sK)V,\qquad j=1,...,\rnk(\Sigma),
\end{equation}
provide a Kraus decomposition for the map $\map{E}$, which then is a
QO. This proves the sufficient condition.  \par For the necessary
condition, consider a QO $\map{E}$ where $\{E_i\}\subset\Bnd{H,K}$ are
the elements of any Kraus decomposition.  From lemma \ref{para}, there
exists a set of operators $\{ F_j\} \subset\Bnd{H,K}$ such that
$\sum_j F_j^\dag F_j= I_{\sH}-\sum_i E_i^{\dag}E_i\ge 0$. Now,
consider a set of orthonormal vectors $\{|e_i\>,|f_j\>\}$ in $\sL$,
and define the orthogonal projector $\Sigma=\sum_i|e_i\>\<e_i |$ and
the isometry 
\begin{eqnarray}
V=\sum_i|e_i\>\otimes E_i +\sum_j|f_j\>\otimes F_j\;.
\label{star}
\end{eqnarray}
These operators will provide the desired dilation in Eq.
(\ref{iStines}).\par To complete the proof we need to show that
$\Sigma=I_\sL$ if and only if $\map{E}$ is trace-preserving. If
$\map{E}$ is trace-preserving, we don't need operators $\{F_j\}$
and hence we can choose the space $\sL$ to be spanned by the vectors
$\{|e_i\>\}$, that form an orthonormal basis for $\sL$, namely
$\sum_i|e_i\>\<e_i|=I_\sL$. On the other hand, if $\Sigma=I_\sL$, one
has $\dual{\map{E}}(I_\sK)=V^\dag V=I_\sH$, namely the map is
trace-preserving.  \qed

Theorem \ref{tuno} allows to derive a bound for the physical resources
that one needs to obtain the dilation (\ref{iStines}) of a QO (as we will
see in subsection III.A, the unitary dilation of the isometry does not
introduce any additional ancillary resource). It is clear that for
trace-preserving maps one has $F_j=0$ for all $j$ in the proof of the
theorem. Notice also that since $V\in\Bnd{H,L\otimes K}$ is
an isometry, one has $\dim(\sH)\le\dim(\sL)\times\dim(\sK)$.  The
minimum dimension for $\sL$ is obtained for the canonical Kraus
decomposition and for the minimum cardinality of the complementary set
of operators $\{F_j\}$ that is given by
$\lceil\rnk(I_\sH-K)/\dim(\sK)\rceil$.  Therefore, upon denoting by
$c$ the cardinality of the canonical Kraus decomposition, namely the
\emph{rank} of the QO, one has
\begin{equation}\label{stricter1}
\left(c+\left\lceil\frac{\rnk(I_\sH-K)}{\dim(\sK)}
\right\rceil\right)\times\dim(\sK)\ge\dim(\sH)
\end{equation}
for every map $\map{E}:\T{H}\rightarrow\T{K}$. In fact, using Lemma 1,
in Eq. (\ref{star}) one has at least $c$ elements $E_i$ and at least
$\lceil r/\dim(\sK ) \rceil$ elements $F_j$, where $r$ is the rank of
$I_{\sH}-K$. 
From Eq.  (\ref{stricter1}) and the condition
$\dual{\map{E}}(I_\sK)=K$, we have
\begin{equation}\label{stricter}
\dim(\sL)\ge
c+\left\lceil\frac{\rnk(I_\sH-\dual{\map{E}}(I_\sK))}{\dim(\sK)}\right\rceil\ge\frac{\dim(\sH)}{\dim(\sK)}\,.
\end{equation}
Eq. (\ref{stricter}) provides a bound on the resources that 
one needs to obtain an isometric dilation, without
knowing \emph{a priori} a Kraus decomposition for the map.
\par In Theorem \ref{tuno} we have shown how to obtain a QO via an
isometric embedding. In the following subsections, we explicitly
derive the physical realizations for the QO for both the free and the
interacting formulations.
\subsection{Free dilations.}
We start by giving the proof of the well known lemma of Gram-Schmidt
unitary dilations \cite{lang}.
\begin{lemma} Every isometry
$T\in\set{B}(\sH_{\rm in},\sH_{\rm out})$ admits a unitary
dilation $U\in\set{B}(\sH_{\rm out})$.
\end{lemma}
\Proof Introduce a Hilbert space $\sH_{\rm aux}$ such that $\sH_{\rm
  out}=\sH_{\rm in}\oplus\sH_{\rm aux}$.  We consider the case
$\dim(\sH_{\rm aux})\ge 1$, otherwise $\sH_{\rm out}\cong\sH_{\rm
  in}$, and $T$ is already unitary. For a given isometry
$W\in\set{B}(\sH_{\rm aux},\sH_{\rm out})$, define the operator
$U\in\set{B}(\sH_{\rm out})$ as
\begin{equation}
\begin{split}
&U=T\hplus W\\ 
&U|v_{\rm out}\>=U(|v_{\rm in}\>\oplus|v_{\rm aux}\>)=T|v_{\rm in}\>+ 
W|v_{\rm aux}\>\;.
\end{split}
\end{equation}
In finite dimension, this can be obtained on a chosen basis just by
joining \emph{horizontally} the two matrices $T$ and $W$ so that, by
construction, $U$ is a square matrix, whence the symbol $\hplus $. If
the condition
\begin{equation}
T^\dag W=0
\label{19}
\end{equation}
is satisfied, then the operator $U$ is unitary on $\sH_{\rm out}$.  An
operator $W\in\Bnd{H_{\rm aux} ,H_{\rm out}}$ that
satisfies Eq. (\ref{19}) has column vectors $[W(k)]$ for
$k=1,\ldots,\dim(\sH_{\rm aux})$ that make an orthonormal basis for
$\sH_{\rm aux}=\Rng(I_{\sH_{\rm out}}- TT^\dag)\subset\sH_{\rm out}$.
A set of vectors of this kind can always be obtained iteratively by
the Gram-Schmidt procedure on $\sH_{\rm in}\oplus\sH_{\rm aux}$, with
$\dim(\sH_{\rm aux})>0$.  \qed
\par Using the previous lemma, one can obtain 
a unitary operator $U\in\Bnd{L\otimes K}$ from the isometry
$V\in\Bnd{H,L\otimes K)}$ by sticking horizontally $V$ with an
appropriate isometry $W\in\Bnd{D,L\otimes K)}$, where $\sD$ is a
second auxiliary Hilbert space defined by the relation
$\sH\oplus\sD=\sL\otimes\sK$. From $U$, reversely, it is possible to
reconstruct $V$ using the \emph{dilation operator}
$D\in\Bnd{H,L\otimes K}$, which is the trivial isometry
$D=I_{\sH}\vplus\mathbf{0}_{\sH,\sD}$ (in analogy with $\hplus$, the
symbol $\vplus$ means the \emph{vertical} joining of two block
matrices, whereas the symbol $\mathbf{0}_{\sH,\sD}$ denotes the
operator in $\Bnd{H,D}$ corresponding to the rectangular matrix with
all zero entries), such that
\begin{equation}
V=UD\;,
\end{equation}
where $D$ acts as follows
\begin{equation}
D|v_{\sH}\>=I_{\sH}|v_{\sH}\>\oplus\mathbf{0}_{\sH,\sD}|v_{\sH}\>=|v_{\sH}\>\oplus
|0\>_\sD\;,
\end{equation}
where $|0\>_\sD$ denotes the null vector in $\sD$.
\par In this way, we can re-express Theorem \ref{tuno} stating that 
\emph{a linear map $\map{E}:\T{H}\rightarrow\T{K}$ is a QO if and
  only if its dual form can be written as}
\begin{equation}\label{unitary}
\dual{\map{E}}(X)=D^\dag U^\dag (\Sigma\otimes X) UD.
\end{equation}
\par Therefore, any trace-decreasing QO can be interpreted in terms of
a unitary interaction between the quantum system and an ancilla,
followed by an orthogonal projection. The dilation operator $D$ is
needed just in order to reduce the output space of the unitary
operator to the original output space of the map.  \par The
Schr\"odinger form of Eq. (\ref{unitary}) can be obtained as
follows. From the duality relation in Eq. (\ref{dual}), one has
\begin{equation}
\begin{split}
\Tr[\dual{\map{E}}(X)\rho] & =\Tr[D^\dag U^\dag (\Sigma\otimes X)UD\ 
\rho]\\
& =\Tr[(\Sigma\otimes X)\ UD\rho D^\dag U^\dag]\\
& =\Tr[X\ \Tr_{\sL}[(\Sigma\otimes I_\sK)
\ UD\rho D^\dag U^\dag ]\ ]\;,
\end{split}
\end{equation}
whence 
\begin{equation}\label{lighterform}
\begin{split}
\map{E}(\rho) & =\Tr_{\sL}[(\Sigma\otimes I_\sK)
\ UD\rho D^\dag U^\dag ]\\
& =\Tr_{\sL}[(\Sigma\otimes I_\sK)
\ U(\rho\oplus\mathbf{0}_{\sD}) U^\dag ]\;,
\end{split}
\end{equation}
where $\mathbf{0}_\sD$ is the null operator on $\sD$. In Eq.
(\ref{lighterform}) the term $U(\rho\oplus\mathbf{0}_\sD)U^\dag$ represents a
\emph{free unitary evolution} of the system in the state $D\rho D^\dag
\equiv\rho\oplus\mathbf{0}_{\sD}$, which is a positive block-diagonal
operator in $\T{L\otimes K}$ with unit trace (remember that
$\sH\oplus\sD=\sL\otimes\sK$). Physically such trivial 
embedding of $\sH$ in $\sH\oplus\sD$ can be regarded as
kind of conservation law or super-selection rule forbidding a subspace
for the input states. 
\par In conclusion of this subsection, we notice
that the special case of $V$ already unitary in Eq. (\ref{iStines})
corresponds to no subspace $\sD$, and $D\equiv I_\sH$ and $U\equiv
V$. Then, Eq. (\ref{lighterform}) becomes simply 
\begin{equation}
\map{E}(\rho)=\Tr_\sL[(\Sigma\otimes I_\sK)\ U\rho U^\dag]\;,
\end{equation}
and one necessarily has $\dim(\sK)\le\dim(\sH)$.

\subsection{Interacting dilations.}
In the previous subsection we have derived a general unitary
realization for a given QO, in terms of a direct-sum dilation, using a
measurement ancilla only, with the input space embedded in a larger
Hilbert space, where a kind of super-selection rule forces the choice
of the input state in a proper subspace before a free unitary
evolution on the extended space. We are now interested in the
tensor-product types of realization schemes, in which the role of the
dilation operator $D$ (i.e. of the super-selection rule) will be
played by the tensor product of $\rho$ with the state of a preparation
ancilla, with the system interacting with such ancilla, and with a
conventional observable measurement then performed on a different
ancilla. This ancilla-system interaction scenario is more popular in
the literature, and is the one used in the extension theorems for
instruments in Refs. \cite{Ozawa84}. It is obvious that also composed
schemes are possible, with both direct-sum and tensor-product
dilations.

The results of the previous subsection can be rewritten by choosing a
dilation in terms of a Hilbert space $\sL$ with dimension
$\dim(\sL)\times\dim(\sK)=r\dim(\sH)$, for 
integer $r$. Then, upon introducing a second ancillary space $\sR$
with $\dim(\sR)=r$, one has $\sL\otimes\sK\cong\sR\otimes\sH$, and we
obtain the following theorem. 
\begin{theorem} \label{daygum}
A linear map $\map{E}:\T{\sH}\longrightarrow\T{\sK}$ is a QO if
and only if its dual form can be written as follows
\begin{equation}
\dual{\map{E}}(X)=\<\phi_\sR|U^\dag (\Sigma\otimes X) U|\phi_\sR\>\;,
\end{equation}
where $X\in\Bnd{K}$ is the input, $\Sigma\in\Bnd{L}$ is
a non-vanishing orthogonal projector on a subspace of the ancillary space
$\sL$, $U\in\Bnd{L\otimes K}$ is unitary and $|\phi_\sR\>\in\sR$
is a fixed normalized vector. In the Schr\"odinger picture one has 
\begin{equation} \label{partelem}
\map{E}(\rho)=\Tr_\sL[(\Sigma\otimes I_\sK)\
U(|\phi_\sR\>\<\phi_\sR|\otimes\rho)U^\dag]\;,
\end{equation}
where now the input is represented by the state
$\rho\in\T{H}$. We refer to the spaces $\sR$ and $\sL$ as
the preparation and measurement ancilla, respectively.
\end{theorem}
\medskip\par\noindent{\bf Proof. } Notice that the notation
$\<\phi_{\set{R}}|U^\dag (\Sigma\otimes X) U |\phi_{\set{R}}\>$
denotes a partial matrix element: in our tensor notation this
corresponds to write $(\<\phi_\sR|\otimes I_\sH)U^\dag(\Sigma\otimes
X)U (|\phi_\sR\>\otimes I_\sH)$.
\par Let us consider the unitary dilation in Eq. (\ref{unitary}), and 
expand the space $\sL$ such that 
\begin{equation}
\dim(\sL\otimes\sK)=\dim(\sL)\times\dim(\sK)=r\dim(\sH)\;,
\end{equation}
for integer $r$. The Hilbert space $\sD$ defined as
$\sD=(\sL\otimes\sK)\ominus\sH$, now has dimension
$\dim(\sD)=\dim(\sL)\times\dim(\sK)-\dim(\sH)=(r-1)\dim(\sH)$.  Let us
introduce a Hilbert space $\sR$ with dimension $\dim(\sR)=r$, so that
\begin{equation}
\dim(\sL\otimes\sK)=\dim(\sR\otimes\sK)\;,
\end{equation}
whence 
\begin{equation}
\sL\otimes\sK\cong\sR\otimes\sH \;.
\end{equation}
Clearly, if the map has equal input and output spaces, the preparation
ancilla and the measurement ancilla are isomorphic, i.e.  $\sR\cong\sL$.
On fixed orthonormal bases for $\sK$, $\sL$, $\sH$ and $\sR$, upon
denoting by $|\phi_\sR\>\in\sR$ an element of the basis of $\sR$,  
we have the following identifications
\begin{equation}
D\equiv|\phi_{\set{R}}\>\otimes I_\sH\;,
\label{r1}
\end{equation}
and
\begin{equation}
\rho\oplus\mathbf{0}_{\set{D}}\equiv
|\phi_{\set{R}}\>\<\phi_{\set{R}}|\otimes\rho\;.
\label{r2}
\end{equation}
The statement of the theorem is then obtained by rewriting Eqs.
(\ref{unitary}) and (\ref{lighterform}) with the use of Eqs.
(\ref{r1}) and (\ref{r2}).  \qed \medskip\par\noindent{\bf Alternative
  proof. }The above proof is based on the direct-sum dilation of Eq.
(\ref{unitary}). An equivalent way to obtain the result in Theorem
\ref{daygum} is the following. From the Stinespring form in Eq.
(\ref{iStines}), let us introduce a Hilbert space $\sR$ such that
$\sL\otimes\sK\cong\sR\otimes\sH$. By a repeated use of the
Gram-Schmidt procedure one obtains other isometries
$W_i\in\Bnd{H,L\otimes K}$, for $i=2,\dots,r$, such that
\begin{equation}
V^\dag W_i=0\quad\textrm{and}\quad W^\dag_iW_j=\delta_{ij}\ I_\sH\;,
\end{equation}
namely
\begin{equation}
\Rng(V)\oplus\Rng(W_2)\oplus\dots\oplus\Rng(W_{r})=\sL\otimes\sK \;.
\end{equation}
Let us consider the unitary operator 
\begin{equation}
U=\<r_1|\otimes V+\<r_2|\otimes W_2+\dots+\<r_r|\otimes W_r\;,
\end{equation}
where $\{|r_i\>\}\subset\sR$ is an orthonormal basis for the space
$\sR$. By taking $|r_1\>\equiv|\phi_\sR\>$, one obtains the statement
of the theorem. \qed This constructive proof has been used in Ref.
\cite{transp} to explicitly derive a unitary realization for the
optimal transposition map.
\par The tensor-product form of the unitary dilation is generally more
expensive in terms of resources (i.e. the dimension of the extended
space) than the direct-sum form in Eq. (\ref{lighterform}), however,
the physical realization of the tensor-product could be more
practical, since one just needs to prepare a fixed ancilla state,
without the need of a super-selection rule.
\par By a further enlargement of the ancilla space, the structure of the
unitary interaction that realizes a given QO can be simplified. The
following derivation generalizes the Halmos method \cite{Halmos}, and
has been used in  \cite{Max} to provide unitary realizations of the ideal phase
measurement. 
\par From the Stinespring dilation (\ref{iStines}), where we take $\sL$ such that 
$\sL\otimes\sK\cong\sR\otimes\sH$, let us define the operators
\begin{equation}
\tilde{V}=V(\<\phi_\sR|\otimes I_\sH)\quad\textrm{and}\quad
\tilde{V}^\dag=(|\phi_\sR\>\otimes I_\sH)V^\dag\;.
\end{equation}
One can simply verify that both $\tilde{V}\tilde{V}^\dag$ and
$\tilde{V}^\dag\tilde{V}$ are projectors, i.e. 
$(\tilde{V}\tilde{V}^\dag)(\tilde{V}\tilde{V}^\dag)=\tilde{V}\tilde{V}^\dag$
and $(\tilde{V}^\dag\tilde{V})(\tilde{V}^\dag\tilde{V})=\tilde{V}^\dag
\tilde{V}$. Let us introduce a \emph{third} ancilla space $\sS$ and a
linear operator $W$ on $\sS$ such that
\begin{equation}\label{condW}
W^2=W^{\dag 2}=0\quad\textrm{and}\quad WW^\dag+W^\dag W=I_\sS\;.
\end{equation}
These conditions imply that $WW^\dag$ and $W^\dag W$ are
orthogonal projectors. We can now write the unitary operator
$U\in\Bnd{S\otimes L\otimes K}$ as follows 
\begin{eqnarray}
U&=&WW^\dag\otimes\tilde{V}-W^\dag W\otimes\tilde{V}^\dag
\nonumber \\&  + &
W^\dag\otimes(I-\tilde{V}^\dag\tilde{V})
+W\otimes(I-\tilde{V}\tilde{V}^\dag)\;, 
\end{eqnarray}
thus obtaining  the map by the equation
\begin{equation}\label{longerform}
\mathcal{E}(\rho)=\Tr_{\sS,\sL}[(I_\sS\otimes\Sigma\otimes I_\sK)\
U(\sigma_\sS\otimes|\phi_\sR\>\<\phi_\sR|\otimes\rho)U^\dag]\;,
\end{equation}
where $\sigma_\sS=\frac{WW^\dag}{\Tr[WW^\dag]}$ is the fixed
normalized state of the third ancilla. Notice that the space $\sS$ and
the operator $W$ are arbitrary, provided that the constraints in Eq. 
(\ref{condW}) are satisfied. For $\dim(\sS)=2$ and $W=|0\>\<1|$ one
recovers the Halmos unitary dilations \cite{Halmos}.

\subsection{Power interacting dilations.}
We have shown in Theorem \ref{daygum} how to obtain a unitary
interaction $U$ that realizes a given QO
$\map{E}:\T{H}\rightarrow\T{K}$, as in Eq. (\ref{partelem}).
Consider now a trace-preserving QO with $\sH\equiv\sK$, namely a
customary {\em channel}. The equivalence of the input and output
spaces implies, in the interacting scheme, the coincidence also
between the preparation and the measurement ancilla, namely
$\sR\equiv\sL$ \cite{note2}. The map $\map{E}$ can now be applied recursively,
and we study the properties of its \emph{powers}
\begin{equation}
\rho\mapsto\map{E}(\rho)\mapsto\map{E}(\map{E}(\rho))\doteq
\map{E}^2(\rho)\mapsto\dots\mapsto\map{E}^n(\rho)\;.
\end{equation} \par
Of course, the unitary realization given in Eq. (\ref{partelem}) does
not satisfy the composition law for powers of the map, namely  
\begin{equation}
\map{E}^n(\rho)\neq\Tr_\sR[U^n(|\phi_\sR\>\<\phi_\sR|\otimes\rho)(U^\dag)^n]\;.
\end{equation}
In fact, the unitary dilation needs a \emph{fresh resource}, i.e. a
{\em disentangled input ancilla}, whereas generally it returns an 
\emph{entangled} output. For this reason, powers of $U$ do not
correspond to powers of $\map{E}$.

\par Here we address the problem of finding \emph{unitary power
  interacting dilations} for a given map.  Using the unitary $U$ and
  the ancilla state $|\phi _\sR \rangle $ of Eq.  (\ref{partelem}),
  let us define the $n$-copy ancilla state
  $\sigma=|\phi_\sR\>\<\phi_\sR|^{\otimes n}$ and the unitary operator
  on $\sR^{\otimes n}\otimes\sH$
\begin{equation}
W=(\Pi_{i=1}^{n-1}E_{i,n}\otimes I_\sH)(I_{1,2,\dots,n-1}\otimes U)\;,
\end{equation}
where the product of swap operators $E_{i,n}|\psi \rangle |\phi
\rangle =|\phi \rangle |\psi \rangle $ for $|\psi \rangle \in \sR _i$
and $|\phi \rangle \in \sR _n$ performs a cyclic permutation of the
ancilla spaces $\sR _i$. One has
\begin{equation}\label{composing}
\map{E}(\rho)=\Tr_{\sR^{\otimes n}}[W\, (\sigma\otimes\rho )\, W^\dag]\;.
\end{equation}
It is now easy to check that the unitary realization in Eq. (\ref{composing})
satisfies the composition law for $k$-powers up to $n$
\begin{equation}
\mathcal{E}^k(\rho)=\Tr_{\sR ^{\otimes n}}
[W^k(\sigma\otimes\rho)(W^\dag)^k]\;,\quad 
k=1,\dots,n\;.
\end{equation}
In fact, the permutation operator selects one \emph{fresh ancilla} at 
\emph{every step} of the interaction, leaving the others unchanged.
\section{Majorization selection of unitary dilations}
We give now a criterion to select the unitary dilations of  a QO 
in terms of a \emph{majorization} relation. We recall that for  
two vectors $x,y\in\mathbb{R}^n$ we say that $x$ is
\emph{majorized} by $y$, i.e.  $x\prec y$, iff 
\cite{Bhatia,Marshall}
\begin{equation}
\sum_{j=1}^k x_j^\downarrow\le\sum_{j=1}^k y_j^\downarrow \;,\qquad 1\le k<n\;,
\end{equation}
and
\begin{equation}
\sum_{j=1}^n x_j^\downarrow=\sum_{j=1}^n y_j^\downarrow\;,
\end{equation}
where $v^\downarrow$ denotes the vector obtained from $v$ by rearranging its
entries in not-increasing order.
\par In Ref. \cite{Niel2000} Nielsen proved the following theorem
that characterizes the ensembles corresponding to a given density
operator $\rho$ by means of a majorization relation.
\par \emph{Let $\rho\in\T{H}$  and
  $(p_i)$ a probability vector. There exist normalized vectors
  $|\psi_i\>\in\sH$ such that
\begin{equation}
\rho=\sum_i p_i |\psi_i\>\<\psi_i|
\end{equation}
if and only if
\begin{equation}\label{Niel}
(p_i)\prec(\lambda_{\rho})\;,
\end{equation}
where $(\lambda_{\rho})$ is the vector of eigenvalues of $\rho$.}
\par We now apply the Nielsen's theorem in order to select the unitary
dilations of a given QO, by exploiting the isomorphism \cite{jami}
between CP maps from $\T{H}$ to $\T{K}$ and positive operators on
$\sK\otimes\sH$.  This correspondence is defined by the relations
\cite{clon,max2001}
\begin{equation}
\begin{split}\label{operatormap}
&R_{\map{E}}=\map{E\otimes I}(|I\kk\bb I|)\;, \\
&\map{E}(\rho )=\Tr _{\sH}[(I_{\sK}\otimes \rho ^T) R_{\map E}]\;,
\end{split}
\end{equation}
where $|I\kk\in\set{H\otimes H}=\sum _n |n \rangle \otimes |n \rangle$
is the maximally entangled unnormalized vector, $T$ denotes the
transposition on the basis $\{|n \rangle \}$,
$\map{I}:\T{H}\rightarrow\T{H}$ is the identity map, and we used
the notation \cite{pla}
\begin{eqnarray}
|A \kk =\sum_{n,m}A_{nm}|n \rangle \otimes |m \rangle \;,\label{}
\end{eqnarray} 
for bipartite pure states. In terms of the positive operator
$R_\map{E}$, the identity (\ref{sum}) becomes $\Tr_\sK[R_\map{E}]=K$. \par
Denoting by $\|A\|_2=\sqrt{\Tr[A^\dag A]}$ the Hilbert-Schmidt norm of
the operator $A$, we have the following theorem.
\begin{theorem}
  Let $\map{E}$ be a QO from $\T{H}$ to $\T{K}$ with canonical
  Kraus decomposition given by $\map{E}(\rho)=\sum_{j=1}^c E_j\rho
  E_j^\dag$ with $\Tr[E_i^\dag E_j]=\|E_i\|^2_2\ \delta_{ij}$. Then
  all the possible unitary interacting dilations for $\map{E}$
  obtained by Theorem \ref{daygum} must satisfy the majorization
  constraint
\begin{equation}\label{maj}
(\|\<\sigma_i|U|\phi_{\sR}\>\|_2^2)
\prec(\|E_i\|^2_2)\;,
\end{equation}
where $\{|\sigma_i\>\}\subset\sL$ form an orthonormal basis for
$\Rng(\Sigma)$ (see Theorem \ref{daygum}).
\end{theorem}
\Proof When representing the CP-map $\map{E}$ with the positive
operator $R_\map{E}$ as in Eq. (\ref{operatormap}), a Kraus
decomposition $\map{E}(\rho)=\sum_i E'_i\rho E'^\dag_i$ for $\map{E}$
can be regarded as the ``ensemble'' realization $R_\map{E}=\sum_i
|E'_i\kk\bb E'_i|$ for the ``density operator'' $R_\map{E}$. Hence,
different Kraus decompositions $\{E'_1,...,E'_m\}$ for $\map{E}$
correspond to different ensembles, with probability vector
$(\|E'_i\|^2_2)$ given by the Hilbert-Schmidt norms of the operators
$E'_i$. On the other hand, the probability vector $(\|E_i\|^2_2)$ of
the canonical Kraus decomposition corresponds to the vector of
eigenvalues of $R_\map{E}$, whence Eq. (\ref{Niel}) in the present
context becomes
\begin{equation}
(\|E'_i\|^2_2)\prec(\|E_i\|^2_2)\;,
\end{equation}
and Eq. (\ref{sum}) guarantees that the two vectors have the same
length.  Then the statement of the theorem follows from the identification
$E'_i=(\<\sigma_i|\otimes I_\sK)U(|\phi_{\sR}\>\otimes I_\sH)$.\qed The
above theorem provides another bound on the dimension of the ancilla
space $\sL$.  Since in Eq.  (\ref{maj}) one has
$i=1,...,\rnk(\Sigma)$, and $j=1,...,c=\rnk(R_{\map{E}})$ ($c$ is the
cardinality of the canonical Kraus decomposition), then
\begin{equation}\label{non-strict}
\dim(\sL)\ge\rnk(\Sigma)\ge c\;.
\end{equation}
This bound can be compared with the tighter one in Eq.
(\ref{stricter}). \par Eq. (\ref{maj}) can also be used to introduce a
partial ordering \cite{note-maj} between all possible unitary
interacting dilations (\ref{partelem}) for the same QO $\map{E}$. In
fact, Eq. (\ref{maj}) states that the unitary interactions from a
canonical Kraus decomposition majorize in the sense of Eq. (\ref{maj})
all those derived from a generic Kraus decomposition. In other words,
the more the Kraus decomposition $\{E'_i\}$ is ``mixed'', i.e. it is an
isometric combination of the canonical one $E'_i=\sum_{j=1}^cY_{ij}E_j$
for $Y^\dag Y=I_c$, the more the unitary interaction constructed with
the $\{E'_i\}$ will be ``flat'' in the Hilbert-Schmidt norms of its
partial matrix elements $\|\<\sigma_i|U|\phi_{\sR}\>\|_2^2$.  This
means that the partial ordering would also reflect a minimization of
the ancillary resource in terms of its Hilbert space dimension.
\section{Conclusions}
Given a QO, generally trace-non-increasing and with different input
and output spaces, we have seen how to obtain its unitary realizations
in terms of both free and interacting dilations. These
different forms of dilation require different amounts of resources in
order to achieve the unitary interaction, and the minimum resource in
terms of Hilbert space dimension is obtained with the free dilation,
where the input state is embedded in a larger Hilbert space and 
a kind of super-selection rule forces the choice of the input state in
a proper subspace before the free unitary evolution. For this case we
derived bounds for the physical resources needed to achieve a QO, in
terms of the dimension of the measurement ancilla space. The
interacting dilations, on the other hand, correspond to the customary
realization in terms of an ancilla-system interaction. Then we have
seen how the construction can be generalized in order to include also
unitary power dilations of a given QO, namely unitary interacting
realizations that also provide the $k$-th power of the map. Finally,
we have seen how all possible interactions can be pre-selected by
means of a majorization inequality, involving the unitary operator, the
ancilla preparation state, and the measured observable.
\section*{Acknowledgments}
This work has been sponsored by INFM through the project
PRA-2002-CLON, and by EEC and MIUR through the cosponsored ATESIT 
project IST-2000-29681 and Cofinanziamento 2002. 
G. M. D. also acknowledges partial support from Department 
of Defense Multidisciplinary University Research Initiative (MURI)
program administered by the Army Research Office under Grant 
No. DAAD19-00-1-0177.
 

\begin{thebibliography}{99}
\bibitem{kron}The usual Kronecker matrix representation uses the following
lexicographic ordering for the basis of the tensor-product Hilbert space: 
the basis element $|i \rangle \otimes |j \rangle  $ precedes $|k
\rangle \otimes |l \rangle  $ if and only if  either $i< k$ or $i=k$
and $j<l$. For example, for $A\otimes B$ one writes 
\begin{equation}
A\otimes B=
\begin{pmatrix}
a_{11}B & a_{12}B & ... & a_{1n}B & ...\\
a_{21}B & a_{22}B & ... & a_{2n}B & ...\\
\vdots & \vdots & \ddots & \vdots & \vdots\\
a_{m1}B & a_{m2}B & ... & a_{mn}B & ...\\
\vdots & \vdots & \vdots & \vdots & \ddots \nonumber 
\end{pmatrix}\;.
\end{equation}
\bibitem{note2} Notice that the isomorphism $\sR\equiv\sL$ of the two
  ancillary Hilbert spaces doesn't mean that the two ancilla are physically
  equivalent, but only that there exists unitary realizations with the
  same physical ancilla. In fact, the ancillary spaces can be embedded
  in different ways in the overall tensor product, and different
  embeddings can be included in the operator of the unitary
  interaction with the system. These kinds of ambiguities are due to
  the implicit identification between physical quantum system and its
  Hilbert space of states within a multipartite (i. e. tensor product)
  composed system. The misidentification is resolved by more properly
  identifying the quantum system with its algebra of observables.
\bibitem{note-maj} Notice that the ordering introduced by a
  majorization relation is neither total nor partial, see Ref.
  \cite{Bhatia,Marshall}. However, defining the equivalence ralation $x\sim y$
  if $y=Px$, for $P$ permutation matrix and $x,y\in\mathbb{R}^n$, it
  can be shown that the majorization ordering $\prec$ is in fact a
  partial ordering on the quotient set $\mathbb{R}^n/\sim$.
\end{thebibliography}
\end{document}